\documentclass[twocolumn,showpacs,preprintnumbers,amsmath,amssymb]{revtex4}

\usepackage{graphicx}
\usepackage{dcolumn}
\usepackage{bm}
\usepackage[english]{babel}
\usepackage{color}
\usepackage{hyperref}
\hypersetup{colorlinks=true,breaklinks,linkcolor=blue,urlcolor=blue,citecolor=blue}

\begin{document}

\preprint{}

\title{Superconductivity, antiferromagnetism and phase separation in\\ the two-dimensional Hubbard model: A dual-fermion approach}

\author{Junya Otsuki$^{1}$}
\author{Hartmut Hafermann$^{2}$}
\author{Alexander I. Lichtenstein$^{3}$}
\affiliation{%
$^1$Department of Physics, Tohoku University, Sendai 980-8578, Japan\\
$^2$Institut de Physique Th\'eorique (IPhT), CEA, CNRS, 91191 Gif-sur-Yvette, France\\
$^3$Institute of Theoretical Physics, University of Hamburg, 20355 Hamburg, Germany
}%

\date{\today}

\begin{abstract}
The dual-fermion approach offers a way to perform diagrammatic expansion around the dynamical mean-field theory.
Using this formalism, the influence of antiferromagnetic fluctuations on the self-energy is taken into account through ladder-type diagrams in the particle-hole channel.
The resulting phase diagram for the (quasi-)two-dimensional Hubbard model exhibits antiferromagnetism and d-wave superconductivity. Furthermore, a uniform charge instability, i.e., phase separation, is obtained in the low doping regime around the Mott insulator. We also examine spin/charge density wave fluctuations including d-wave symmetry. The model exhibits a tendency towards an unconventional charge density-wave, but no divergence of the susceptibility is found.
\end{abstract}


\maketitle

\section{Introduction}
Magnetism and superconductivity appear nearby in typical phase diagrams of transition-metal and heavy-fermion compounds.
Magnetism is related to the Mott insulating state and heavy-fermion formation, which can be described in terms of local correlations.
On the other hand, unconventional superconductivity requires spatial correlations to be taken into account.
For a comprehensive understanding, therefore, one needs a unified treatment of local correlations and spatial fluctuations, which has been a theoretical challenge in the field of strongly correlated electron systems.

A long-standing problem, which may be related to magnetism and superconductivity, is the pseudo-gap state in the low-doped regime of cuprates~\cite{Timusk99}. One of the candidates for its origin is a hidden order, i.e. the staggered flux state or the d-density wave (d-DW)~\cite{Chakravarty01}.
There are some experiments which indicate broken time-reversal symmetry in the pseudo-gap regime~\cite{Fauque06, Shekhter13}. 
Theoretically, the mean-field approximation based on the slave-boson representation yields a d-DW in the $t$-$J$ model~\cite{Kotliar88,Ubbens92,Wen96}.
However, no clear evidence for the transition has been found in the Hubbard model~\cite{Honerkamp02, Macridin04, Lu12, Yokoyama-commun}.

Another feature, which possibly emerges near the Mott insulator, is a uniform charge instability, i.e., phase separation between two states with different electron density. It was pointed out for the $t$-$J$ model on the basis of energy arguments~\cite{Emery90}, and was indeed demonstrated numerically in the one-dimensional system~\cite{Ogata91} and in infinite dimensions~\cite{Otsuki-Vollhardt}.
In contrast to the d-DW, the phase separation has been observed also in the Hubbard model by means of various numerical methods~\cite{Zitzler02,Kotliar02,Capone06,Macridin06,Werner07,Eckstein07,Aichhorn07, Yokoyama13, Misawa-arXiv},
while quantum Monte Carlo investigations reported no evidence of phase separation~\cite{Moreo91,Becca00}.

The unresolved problems described above motivate us to investigate the two-dimensional Hubbard model as a prototypical model of strongly correlated electron systems, and to develop new theories which could clarify these issues.
The dynamical mean-field theory (DMFT) provides a description of the Mott transition~\cite{Georges96} and its cluster extensions provide a route to the d-wave superconductivity (d-SC) in the doped regime~\cite{Maier-RMP, Potthoff03}. 
The d-SC has indeed been obtained in several numerical calculations~\cite{Lichtenstein00, Maier05, Gull13, Capone06}.
We note that cluster DMFT particularly accounts for short-range correlations in addition to the local ones.

A different kind of extension of single-site DMFT has been worked on, which,
in contrast to cluster extensions, aims at incorporating long-range correlations~\cite{Kusunose06, Toschi07, Held08, Katanin09, Slezak09, Taranto-arXiv, Kitatani}. The common idea of these approaches is to introduce an additional step of solving the lattice problem in a certain way after the DMFT equations are solved.
The various formulations differ (i) physically, in the sets of diagrams which are summed beyond DMFT and (ii) technically, how double counting of correlation effects is avoided, that may arise when two different methods are combined.

Rubtsov \textit{et al.} introduced an auxiliary fermion 
which mediates itinerancy of electrons~\cite{Rubtsov08, Rubtsov09}.
With this dual fermion, a perturbation expansion around the DMFT has been made possible
without the double-counting problem; the zeroth-order approximation in this theory corresponds to DMFT, and spatial correlations are systematically incorporated by summing up a series of diagrams. 
In particular, ladder diagrams similar to those in the fluctuation exchange approximation (FLEX)~\cite{FLEX1, FLEX2, Takimoto02, Yanase03} yield descriptions of collective modes (long-range fluctuations). 
Indeed, it has been shown that inclusion of the ladder diagrams in the dual-fermion approach leads to paramagnon excitations that exhibit antiferromagnetic (AFM) fluctuations in the paramagnetic state~\cite{Hafermann09, Hafermann-book}. 
At the same time, the ladder approximation yields suppression of the AFM phase transition in two dimensions~\cite{Hafermann09, Hafermann-book} and the expected critical exponents in case that phase transitions are found~\cite{Antipov14}, 
demonstrating that long-range fluctuations essential for the critical behavior are appropriately included. 
Therefore, the dual-fermion approach with ladder-type diagrams provides a combined description of strong local correlations and long-range correlations.

Although first results of the ladder approximation have been presented in 2009~\cite{Hafermann09,Hafermann09-2}, its exemplary results for doped Mott insulators have been limited because of some technical difficulties arising from strong AFM fluctuations. In this paper, we overcome these limitations and present systematic results for the doped regime of the two-dimensional Hubbard model. We address possible phase transitions of the d-DW and the phase separation in the doped Mott insulator as well as the d-SC.  Our results reveal further characteristics of the ladder approximation.

The rest of this paper is organized as follows.
In the next section, we first present phase diagrams obtained in this investigation to give an overview of our results.
Afterwards, the dual-fermion formalism and the self-energy equation are presented in Section~\ref{sec:dualfermion}. 
Succeeding Sections~\ref{sec:AFM}--\ref{sec:DW} present detailed numerical results and related formulas for the AFM susceptibility, superconductivity, phase separation, and unconventional density waves.
The paper is closed with discussions in Section~\ref{sec:summary}.

\section{Overview}
\label{sec:overview}

\begin{figure}[tb]
	\begin{center}
	\includegraphics[width=\linewidth]{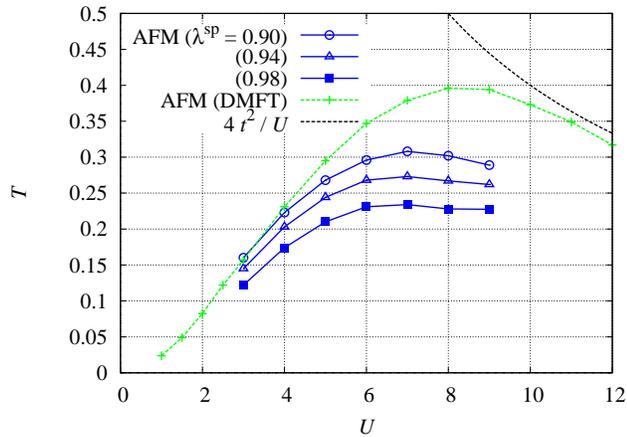}
	\end{center}
	\caption{(Color online) A phase diagram at half filling, $\delta=0$.}
	\label{fig:phase-n1}
\end{figure}

Prior to presenting formalism and detailed numerical results, we first give an overview of our results obtained in this paper.
We investigate the two-dimensional Hubbard model:
\begin{align}
H = \sum_{\bm{k}\sigma} \epsilon_{\bm{k}} c_{\bm{k}\sigma}^{\dag} c_{\bm{k}\sigma}
+ U \sum_{\bm{r}} n_{\bm{r}\uparrow} n_{\bm{r}\downarrow},
\end{align}
with $\epsilon_{\bm{k}}=-2t (\cos k_x + \cos k_y)$.
The number operator $n_{\bm{r}\sigma}$ is defined by
$n_{\bm{r}\sigma}=N^{-1} \sum_{\bm{k}\bm{q}} c_{\bm{k}\sigma}^{\dag} c_{\bm{k}+\bm{q}\sigma} e^{i\bm{q}\cdot \bm{r}}$,
where $N$ denotes the number of lattice sites. We take $t=1$ as the unit of energy.

In two-dimensional systems, the AFM transition is forbidden at $T>0$ by the Mermin-Wagner theorem~\cite{Mermin-Wagner}. This leads to the critical behavior $\chi \sim e^{c \beta}$ of the susceptibility at low temperatures~\cite{Chakravarty88, Hasenfratz91}.
Our approximation indeed shows no AFM transition within calculated temperatures. 
To quantify the AFM fluctuations, we define a ``phase boundary" by the points where the fluctuations exceed a certain criterion (see Section~\ref{sec:AFM} for details). We may regard this line as a phase boundary in quasi-two dimensions. The phase diagram at half filling obtained in this way is shown in Fig.~\ref{fig:phase-n1}.
We plot three phase boundaries corresponding to different criteria. In DMFT, there exists a real phase transition, which is plotted for comparison.

According to a cluster DMFT calculation with a paramagnetic bath~\cite{Park08}, the Mott transition takes place at $U\simeq 6$ and below $T \simeq 0.1$~\cite{footnote-Mott}.
We could not reach this regime due to the critical AFM fluctuations, which renders the self-energy calculation unstable.
We note, however, that cluster DMFT does not take into account critical fluctuations characteristic of two dimensions, meaning that the AFM transition takes place at a higher temperature than the Mott transition. Hence the latter is actually hidden by the AFM phase in cluster DMFT.

\begin{figure}[tb]
	\begin{center}
	\includegraphics[width=\linewidth]{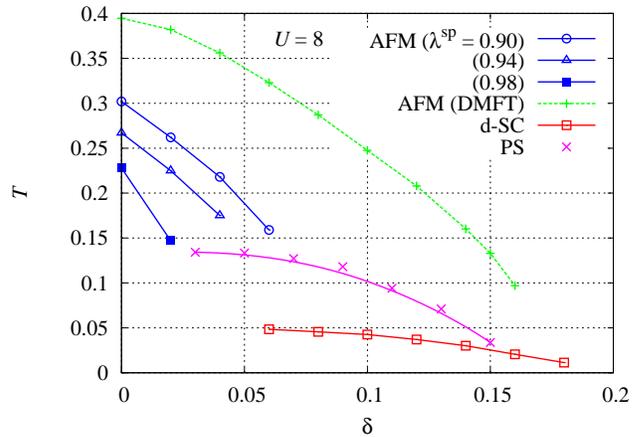}
	\end{center}
	\caption{(Color online) Phase diagrams under doping $\delta=1-n$ for $U=8$.}
	\label{fig:phase-dope}
\end{figure}

Figure~\ref{fig:phase-dope} shows the phase diagram of temperature against doping $\delta=1-n$ for $U=8$.
The d-SC is obtained in the region $T\lesssim 0.05$ and $\delta \lesssim 0.18$.
The superconducting transition temperature $T_{\rm c}$ monotonically increases approaching half filling ($\delta=0$). This behavior is reminiscent of the FLEX~\cite{Takimoto02, Yanase03} and differs from that in cluster DMFT, where the d-SC phase exhibits a maximum at finite doping~\cite{Lichtenstein00, Gull13}.
We consider that the monotonic behavior of $T_{\rm c}$ in our results is due to insufficient treatment of short-range spin fluctuations, which will be discussed in Sec.~\ref{sec:summary}.

In the low-doping regime above $T_{\rm c}$, we found a phase separation. The line $T_{\rm PS}$ in Fig.~\ref{fig:phase-dope} shows the spinodal line, where the uniform charge susceptibility diverges.
The phase separation extends up to $\delta \simeq 0.15$.
At $T<T_{\rm PS}$, the solution is thermodynamically unstable because of $\partial n/\partial \mu < 0$. 
Thermodynamic stability is acquired by inhomogeneous coexistence of regions with different doping levels: Mott insulating regions with $\delta=0$ and metallic regions with larger doping $\delta \neq 0$.
The phase boundary for the d-SC has been computed with the homogeneous solution, which, in fact, is thermodynamically unstable in the region $\delta \lesssim 0.15$. 
Therefore, pure d-SC only realizes in the region $0.15 \lesssim \delta \lesssim 0.18$, while it may not occur for $\delta \lesssim 0.15$.  We have also examined the possibility of a d-DW. We find that the d-DW dominates over DWs with other symmetries, but the corresponding susceptibility shows no divergence.

\section{Dual-ladder approximation}
\label{sec:dualfermion}

\subsection{Dual action}
In the dual-fermion approach, the lattice model is solved in two steps. First, an effective impurity model, which is the same as in DMFT, is solved with the aid of some numerical methods. The local correlations, which are essential for formation of the Mott gap, are fully taken into account at this stage. 
In the next step, an interacting lattice model is constructed by quantities evaluated in the first step, and is solved by a diagrammatic perturbation theory.
This way, spatial fluctuations are incorporated in addition to the local correlations in DMFT.
In the following, we first give a brief summary of the dual-fermion approach~\cite{Rubtsov08, Rubtsov09, Hafermann-book}.

It is convenient to work in the path-integral representation.
The partition function $Z$ is written in terms of Grassmann variables 
$c_{\bm{r}}(\tau)$ and $c_{\bm{r}}^*(\tau)$: 
$Z = \int \prod_{\bm{r}} {\cal D}[c_{\bm{r}}^* c_{\bm{r}}] e^{-{\cal S}[c^*, c]}.$
The action ${\cal S}$ is given by
\begin{align}
{\cal S}[c^*, c]
&= \sum_{\bm{r}} {\cal S}_{\rm imp}[c_{\bm{r}}^*, c_{\bm{r}}] 
+ \sum_{\omega \bm{k} \sigma} (\epsilon_{\bm{k}} -\Delta_{\omega})
 c_{\omega \bm{k} \sigma}^* c_{\omega \bm{k} \sigma},
\label{eq:S_lat}
\end{align}
where the Fourier transform of $c(\tau)$ is defined by 
$c_{\omega}=\beta^{-1/2} \int_0^{\beta} c(\tau) e^{i\omega\tau}$
with $\omega$ being the fermionic Matsubara frequency.
The first term ${\cal S}_{\rm imp}$ describes the effective impurity model of DMFT~\cite{Georges96}:
\begin{align}
{\cal S}_{\rm imp}[c_{\bm{r}}^*, c_{\bm{r}}]
= &- \sum_{\omega \sigma} (i\omega + \mu - \Delta_{\omega})
c^*_{\omega \bm{r} \sigma} c_{\omega \bm{r} \sigma}
\nonumber \\
&+ U \int d\tau n_{\bm{r}\uparrow}(\tau) n_{\bm{r}\downarrow}(\tau).
\label{eq:S_imp}
\end{align}
The hybridization function $\Delta_{\omega}$ is actually canceled out in Eq.~(\ref{eq:S_lat}), 
but an approximate solution may depend on $\Delta_{\omega}$.
A condition for determining $\Delta_{\omega}$ will be discussed later.

In order to construct a lattice model for which the solution of ${\cal S}_{\rm imp}$ is the starting point, Rubtsov \textit{et al.} introduced an auxiliary fermion which ``decouples'' the kinetic-energy term~\cite{Rubtsov08,Rubtsov09} [the second term in Eq.~(\ref{eq:S_lat})].
This fermion is termed dual fermion and represented by $f$.
The dual fermions locally hybridize with the electrons and mediate the electron itinerancy.
The point is that the transformed action written with $c$ and $f$ variables has only local terms concerning $c$ variables.
Therefore, one can integrate out $c$ variables at each site independently. This process corresponds to solving the effective impurity problem expressed by ${\cal S}_{\rm imp}$.
The local hybridization between $c$ and $f$ introduces effective interaction terms among the $f$ variables, which are local in space but non-local in the time domain.

The resulting partition function thus consists only of the dual variables $f$.
Hence, our task now is to solve the dual system described by
$\tilde{Z} =
\int \prod_{\bm{r}} {\cal D}[f_{\bm{r}}^* f_{\bm{r}}]
 e^{-\tilde{\cal S}[f^*, f]}$.
The action $\tilde{\cal S}$ is given by
\begin{align}
\tilde{\cal S}[f^*, f] 
&= -\sum_{\omega \bm{k} \sigma}
 (\tilde{G}^0_{\omega \bm{k}})^{-1} f_{\omega \bm{k} \sigma}^* f_{\omega \bm{k} \sigma}
+ \tilde{V}[f^*, f],
\label{eq:S-dual}
\end{align}
with the bare dual Green's function $\tilde{G}^0_{\omega\bm{k}}$ defined by
\begin{align}
\tilde{G}^0_{\omega \bm{k}}
= (g_{\omega}^{-1} + \Delta_{\omega} - \epsilon_{\bm{k}})^{-1}
- g_{\omega}.
\label{eq:G0_dual}
\end{align}
Here $g_{\omega}=-\langle c_{\omega \bm{r} \sigma} c_{\omega \bm{r} \sigma}^* \rangle_{\rm imp}$ is the impurity Green's function, with $\langle \cdots \rangle_{\rm imp}$ being a thermal average with respect to the action ${\cal S}_{\rm imp}$.
The first term in Eq.~(\ref{eq:G0_dual}) corresponds to the lattice Green's function in DMFT. Subtracting the second term excludes double counting of local correlations.
The term $\tilde{V}$ denotes local interactions, which include many-body interactions as well as a two-body term.
The point of the transformed action $\tilde{\cal S}$ is that the bare propagator $\tilde{G}^0$ and the interaction $\tilde{V}$ fully include local correlations. Hence, the (undressed) dual fermions $f$ may be regarded as particles which involve all the local interaction processes. Residual interactions between the dressed particles are described by $\tilde{V}$.

Once the dual Green's function $\tilde{G}_{\omega\bm{k}}=-\langle f_{\omega\bm{k}\sigma} f_{\omega\bm{k}\sigma}^* \rangle_{\tilde{S}}$ is evaluated, 
it is readily transformed to the Green's function $G_{\omega \bm{k}}$ of the original electrons
by means of the exact relation
\begin{align}
G_{\omega \bm{k}}^{-1} 
&= (g_{\omega} + g_{\omega} \tilde{\Sigma}_{\omega\bm{k}} g_{\omega})^{-1} + \Delta_{\omega} - \epsilon_{\bm{k}}.
\end{align}
Here, we introduced the dual self-energy 
$\tilde{\Sigma}_{\omega\bm{k}}=(\tilde{G}^0_{\omega\bm{k}})^{-1} - \tilde{G}_{\omega\bm{k}}^{-1}$.
It is clear from this expression that $\tilde{\Sigma}_{\omega\bm{k}}=0$ leads to the DMFT formula for the lattice Green's function.

The formalism presented above is still exact.
In the following, two approximations will be made.
Firstly, we retain only two-body interactions in $\tilde{V}$.
Secondly, we perform a perturbation expansion with respect to $\tilde{V}$ to sum up a certain set of diagrams for $\tilde{\Sigma}_{\omega\bm{k}}$. 
These approximations rely on the idea that the DMFT is a good starting point for Mott insulators, and hence the spatial correlations may be dealt with perturbatively.
We can also endorse this treatment by arguments based on the $1/d$ expansion, which will be discussed later.

\subsection{Interaction vertex for dual fermions}
\begin{figure}[tb]
	\begin{center}
	\includegraphics[width=0.95\linewidth]{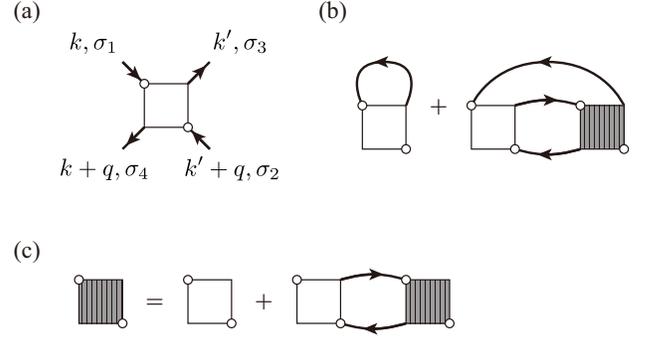}
	\end{center}
	\caption{Diagrammatic representations of (a) bare interaction $\tilde{V}$ for dual fermions, (b) the dual self-energy $\tilde{\Sigma}$ in the ladder approximation, and (c) the equation for the renormalized vertex $\Gamma$.}
	\label{fig:diagrams}
\end{figure}

We retain only two-body interactions in $\tilde{V}$ neglecting terms involving more than three particles. 
Thus, $\tilde{V}$ reads
\begin{align}
\tilde{V}
=-\frac{1}{4} \sum_{kk'q} \sum_{\sigma_1 \sigma_2 \sigma_3 \sigma_4}
\gamma_{\omega\omega'; \nu}^{\sigma_1 \sigma_2 \sigma_3 \sigma_4} 
f_{k\sigma_1}^* f_{k'+q,\sigma_2}^* f_{k'\sigma_3} f_{k+q,\sigma_4},
\label{eq:V_tilde}
\end{align}
where $k=(\omega, \bm{k})$ and $q=(\nu, \bm{q})$ with $\nu$ being the bosonic Matsubara frequency.
The interaction coefficient $\gamma$ corresponds to the vertex evaluated in the effective impurity system. It is defined through
\begin{align}
\langle c_1 c_2 c_3^* c_4^* \rangle_{\rm imp}
&= g_1 g_2 ( \delta_{14} \delta_{23} - \delta_{13} \delta_{24})
+ T g_{1} g_{2} \gamma_{1234} g_{3} g_{4}.
\end{align}
Here, a simplified notation is used such as $1\equiv (\omega_1, \sigma_1)$.
Using energy conservation, we parameterize the frequency dependence of $\gamma$ as
\begin{align}
\gamma_{\omega\omega'; \nu}^{\sigma_1 \sigma_2 \sigma_3 \sigma_4}
&\equiv \gamma_{(\omega, \sigma_1), (\omega'+\nu, \sigma_2), (\omega', \sigma_3), (\omega+\nu, \sigma_4)}.
\end{align}
The antisymmetric nature of $\gamma$ leads to the relation 
$\gamma_{\omega\omega'; \nu}^{\sigma_1 \sigma_2 \sigma_3 \sigma_4}
=-\gamma_{\omega,\omega+\nu; \omega'-\omega}^{\sigma_1 \sigma_2 \sigma_4 \sigma_3}$.
The interaction $\tilde{V}$ is represented by the diagram in Fig.~\ref{fig:diagrams}(a).

We consider the spin dependence of $\gamma$. Using the Pauli matrix $\sigma^{\xi}$ ($\xi=0,x,y,z$, including the unit matrix $\sigma^0$), we transform $\gamma$ as
\begin{align}
&\gamma^{\sigma_1 \sigma_2 \sigma_3 \sigma_4} 
= \frac{1}{2} \sum_{\xi\xi'} \gamma^{\xi\xi'} 
\sigma^{\xi}_{\sigma_1\sigma_4} \sigma^{\xi'}_{\sigma_2\sigma_3}.
\label{eq:gamma-pauli}
\end{align}
Without magnetic field, $\gamma^{\xi\xi'}$ is diagonal and there are only two independent components: 
$\gamma=\text{diag}(\gamma^{\rm ch}, \gamma^{\rm sp}, \gamma^{\rm sp}, \gamma^{\rm sp})$.
By inverting Eq.~(\ref{eq:gamma-pauli}), we obtain
\begin{align}
&\gamma^{\rm ch} \equiv \gamma^{00}
=\frac{1}{2} \sum_{\sigma\sigma'} \gamma^{\sigma\sigma'\sigma'\sigma},
\\
&\gamma^{\rm sp} \equiv \gamma^{zz}
=\frac{1}{2} \sum_{\sigma\sigma'} \sigma\sigma' \gamma^{\sigma\sigma'\sigma'\sigma},
\end{align}
which corresponds to interactions in charge and longitudinal-spin channels, respectively.
The transverse-spin channel
$\gamma^{\rm \perp} \equiv
\gamma^{\uparrow\downarrow\uparrow\downarrow}
=(\gamma^{xx}-i\gamma^{xy})$
is equivalent to $\gamma^{\rm sp}$,
since we are considering the paramagnetic state.

\subsection{Self-energy}
We evaluate the dual self-energy $\tilde{\Sigma}_{\omega\bm{k}}$ 
taking the two-body interaction $\tilde{V}$ in Eq.~(\ref{eq:V_tilde}) into account.
In principle, one can apply any numerical method as well as approximations for this purpose.
Here, we use a perturbation theory and sum up certain diagrams based on physical considerations and a $1/d$ analysis.

We first discuss the choice of diagrams from a physical point of view. 
Near the Mott insulator, the important ingredient are spin fluctuations, which can be taken into account by ladder-type diagrams.
Indeed, the ladder approximation gives a magnetic spectrum which exhibits low-energy spin excitations (so-called paramagnons), as a consequence of strong AFM fluctuations~\cite{Hafermann09, Hafermann-book}. 
In the following, we present the self-energy formula in the ladder approximation, stressing on the SU(2) symmetry for the spin indices.

We first evaluate the renormalized vertex $\Gamma$ collecting successive particle-hole excitations. 
Since the propagator $\tilde{G}$ is independent of the spin component, the vertex $\Gamma^{\alpha}$ for each channel $\alpha=\text{ch}, \text{sp}$ independently obeys the Bethe-Salpeter equation: 
\begin{align}
\Gamma_{\omega\omega';\nu\bm{q}}^{\alpha}
=\gamma_{\omega\omega'; \nu}^{\alpha}
+ T \sum_{\omega''}
\gamma_{\omega\omega''; \nu}^{\alpha}
\tilde{\chi}^0_{\omega''; \nu\bm{q}}
\Gamma_{\omega''\omega'; \nu\bm{q}}^{\alpha},
\label{eq:Gamma}
\end{align}
where $\tilde{\chi}^0$ is defined by
\begin{align}
\tilde{\chi}^0_{\omega; \nu \bm{q}} 
= -\frac{1}{N} \sum_{\bm{k}}
\tilde{G}_{\omega \bm{k}} \tilde{G}_{\omega+\nu, \bm{k}+\bm{q}}.
\end{align}
Figure~\ref{fig:diagrams}(c) shows the diagrammatic representation of the above equation.
The self-energy is evaluated with the renormalized vertex $\Gamma$ as follows:
\begin{align}
\tilde{\Sigma}_{\omega\bm{k}}
&= -\frac{T}{N} \sum_{\omega'\bm{k}'}
 \gamma^{\rm ch}_{\omega\omega'; 0}
\tilde{G}_{\omega'\bm{k}'}
\nonumber \\
&+ \frac{T}{4N} \sum_{\nu\bm{q}}
\tilde{G}_{\omega+\nu, \bm{k}+\bm{q}}
( V^{\rm ch} + 3V^{\rm sp})_{\omega\omega; \nu\bm{q}},
\label{eq:self-ladder}
\end{align}
where the effective interaction $V^{\alpha}$ is defined by
\begin{align}
V^{\alpha}_{\omega\omega'; \nu\bm{q}}
&=
T\sum_{\omega''}
\gamma^{\alpha}_{\omega\omega''; \nu} \tilde{\chi}^0_{\omega'';\nu\bm{q}}
\left[ 2\Gamma^{\alpha}_{\omega''\omega'; \nu\bm{q}} 
 -\gamma^{\alpha}_{\omega''\omega'; \nu} \right].
\label{eq:V_phi}
\end{align}
The corresponding diagram for the dual self-energy is shown in Fig.~\ref{fig:diagrams}(b).

\subsection{A perspective from a $1/d$ expansion}
The momentum dependence of the self-energy disappears in the limit of $d=\infty$ dimensions.
This means that DMFT provides the exact solution of fermionic models in $d=\infty$~\cite{Metzner89, Georges96}.
Since the dual-fermion approach offers an expansion around DMFT, it is reasonable to classify diagrams in terms of $1/d$. In this view, we reconsider the self-energy diagram presented above.

\begin{figure}[tb]
	\begin{center}
	\includegraphics[width=0.95\linewidth]{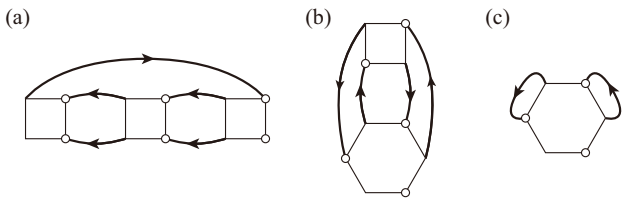}
	\end{center}
	\caption{Self-energy diagrams: (a) Example of a leading order diagram in terms of $1/d$ but not included in the present approximation for physical reasons (see text), (b) an example of the next-leading contributions in the $1/d$ expansion, and (c) zero contribution by the self-consistency condition.}
	\label{fig:diagrams-aux}
\end{figure}

We consider the large-$d$ limit with the scaling $t \propto 1/\sqrt{d}$~\cite{Metzner89, Georges96}.
The local Green's function $G_{\bm{r}=0}$ of the original electrons is of zeroth order in $1/d$, while its dual counterpart vanishes, $\tilde{G}_{\bm{r}=0}=0$, by the self-consistency condition given later (we omit the $\omega$ index for simplicity).
Hence, the dual Green's function has only intersite components $\tilde{G}_{\bm{r}\neq0}$ which scale as $\tilde{G}_{\bm{r}\neq0} \sim G_{\bm{r}\neq0} \sim {\cal O}(1/\sqrt{d})$.
The second-order diagram for $\tilde{\Sigma}_{\omega\bm{k}}$ (the second term in Fig.~\ref{fig:diagrams}(b) with the renormalized vertex replaced by the bare interaction) has a contribution of order ${\cal O}(1/\sqrt{d})$.
The ladder diagrams summed up in Fig.~\ref{fig:diagrams}(c) are of the same order, because the factor $1/d$ arising from the two propagators is canceled by the lattice summation.
Therefore, all diagrams included in the second-term of Fig.~\ref{fig:diagrams}(b) provide the leading contributions of order $1/\sqrt{d}$.
Actually, ladder diagrams in the particle-particle channel also have contributions of the same order (e.g., the diagram in Fig.~\ref{fig:diagrams-aux}(a)). However, we may neglect them since particle-particle fluctuations only have a minor effect in the doped Mott insulator. 
Diagrams containing higher-order vertices only appear at second-to-leading order [e.g., Fig.~\ref{fig:diagrams-aux}(b)], since the local diagram like Fig.~\ref{fig:diagrams-aux}(c) vanishes as explained later.
It means that the three-body and higher-order interactions do not enter to leading order of the $1/d$ expansion.
Indeed, it has been numerically confirmed that the ladder-type diagrams dominate over diagrams built from the three-particle vertex~\cite{Hafermann09}.
In conclusion, the dual-ladder self-energy in Eq.~(\ref{eq:self-ladder}) constitutes the leading correction to the DMFT around the $d=\infty$ limit.

\subsection{Self-consistency condition}
So far, the hybridization function $\Delta_{\omega}$ is arbitrary.
We discuss here how to determine $\Delta_{\omega}$.
The condition is that the scheme should reduce to DMFT if no self-energy corrections are taken into account: $\tilde{\Sigma}_{\omega\bm{k}}=0$.
The following self-consistency condition fulfills this requirement~\cite{Rubtsov08, Rubtsov09, Hafermann-book}:
\begin{align}
\sum_{\bm{k}} \tilde{G}_{\omega\bm{k}} = 0.
\end{align}
It is clear from Eq.~(\ref{eq:G0_dual}) that when $\tilde{\Sigma}_{\omega\bm{k}}=0$, this condition leads to the DMFT self-consistency condition.
Furthermore, it eliminates the contribution from the Hartree diagram (the first term in Fig.~\ref{fig:diagrams}(b)). Similarly, all diagrams which have a propagator connecting the same site (local loops) give no contribution [e.g., Fig.~\ref{fig:diagrams-aux}(c)].

\subsection{Technical details}
We solve the effective impurity problem using the hybridization-expansion solver (CT-HYB)~\cite{Werner06} of the continuous-time quantum Monte Carlo method~\cite{Rubtsov05, Gull-RMP}.
The vertex $\gamma_{\omega\omega'; \nu}$ as well as $g_{\omega}$ are computed. We applied an efficient implementation for the vertex calculation~\cite{Hafermann12}.
The vertex $\gamma_{\omega \omega'; \nu}$ is computed in a small energy window, while the energy cutoff for $g_{\omega}$ can be taken sufficiently large ($10^3$--$10^4$ Matsubara frequency points) in the CT-HYB algorithm. To be concrete, we restrict the frequencies of $\gamma_{\omega \omega'; \nu}$ to $|\omega|, |\omega'| \leq (2n_{\rm c}+1) \pi T$ and $|\nu| \leq 2m_{\rm c} \pi T$.
Typically, we take $n_{\rm c}=m_{\rm c}=20$ for $T \gtrsim 0.1$, and up to $n_{\rm c}=m_{\rm c}=60$ for lower temperatures.
Such a small cutoff compared to the one for $g_{\omega}$ is possible because the frequency summation of $\gamma_{\omega \omega'; \nu}$ is always taken with $\tilde{G}_{\omega\bm{k}}$, which decays faster than the ordinary Green's function, $\tilde{G}_{\omega\bm{k}} \sim -\epsilon_{\bm{k}}/\omega^2$.
We note that the negative bosonic frequencies, $\nu<0$, need not be computed since we have the relation 
$\gamma_{\omega \omega'; \nu}^{\alpha}=(\gamma_{-\omega -\omega'; -\nu}^{\alpha})^*$.

The quantities $g_{\omega}$ and $\gamma_{\omega\omega'; \nu}$ are plugged into the dual-lattice calculations.
The momentum summation (convolution) in Eq.~(\ref{eq:self-ladder}) is evaluated in the real space. Here we can use FFT to reduce ${\cal O}(N^2)$ calculation into ${\cal O}(N\log N)$.
On the other hand, we simply add the frequency $\nu$ in Eq.~(\ref{eq:self-ladder}), since the frequency summation does not simplify considerably in the imaginary-time domain due to the full frequency dependence of $\gamma_{\omega\omega'; \nu}$.
The lattice size $N$ is fixed at $N=32\times32$ (excepting Fig.~\ref{fig:T-afm-scaling}). This size is sufficiently large for our purpose of revealing possible phase transitions. 
A larger system size is necessary to observe the critical behavior, which will be discussed in the next section.
We compute $\tilde{\Sigma}_{\omega\bm{k}}$ and $\tilde{G}_{\omega\bm{k}}$ iteratively until they are converged. To get convergence, we mix new and old data of $\tilde{\Sigma}_{\omega\bm{k}}$. 
The weight of the new data ranges from 0.5 down to 0.02 by checking the tendency toward convergence.

After $\tilde{\Sigma}_{\omega\bm{k}}$ is obtained, we update the bath $\Delta_{\omega}$ and go back to the impurity problem.
We use the formula~\cite{Rubtsov09} 
$\Delta_{\omega}^{\rm new}
=\Delta_{\omega}^{\rm old}
+\xi \tilde{G}_{\omega, \bm{r}=0} / [g_{\omega} ( g_{\omega} + \tilde{G}_{\omega, \bm{r}=0})]$.
Here, $\xi$ is the mixing parameter and typically we take $\xi=0.5$.

When there exist strong AFM fluctuations, i.e., near the half-filling, the iteration for $\Delta_{\omega}$ is unstable.
In this case, we need an elaborate treatment of $\tilde{\Sigma}_{\omega\bm{k}}$ to avoid the numerical instability (see Appendix~\ref{app:stabilize} for details).

\section{Antiferromagnetic Susceptibility}
\label{sec:AFM}
We first present numerical results for the AFM susceptibilities at half filling, $n=1$.
We shall show data for lower temperatures than in the previous calculation of the ladder approximation~\cite{Hafermann09}, and discuss the point of our calculations to achieve convergence in the critical regime.
The results may also be regarded as a benchmark of our calculations.

The spin and charge susceptibilities of the original electrons, $\chi_{\nu\bm{q}}^{\alpha}$, are connected to the reducible vertex part of the dual fermions by an exact relation~\cite{Li08, Brener08, Hafermann-book}.
In the ladder approximation, we use $\Gamma^{\alpha}$ in Eq.~(\ref{eq:Gamma}) for the reducible vertex to obtain explicit expression for $\chi_{\nu\bm{q}}^{\alpha}$ as
\begin{align}
\chi_{\nu\bm{q}}^{\alpha} = \chi_{\nu\bm{q}}^0 
+ T^2 \sum_{\omega\omega'} X_{\omega; \nu\bm{q}} \Gamma_{\omega\omega'; \nu\bm{q}}^{\alpha} X_{\omega'; \nu\bm{q}},
\label{eq:chi_lat}
\end{align}
where
\begin{align}
\chi_{\nu\bm{q}}^0 &= -\frac{T}{N} \sum_{\omega \bm{k}}
G_{\omega\bm{k}} G_{\omega+\nu, \bm{k}+\bm{q}},
\\
X_{\omega; \nu\bm{q}} &= -\frac{1}{N} \sum_{\bm{k}}
G_{\omega\bm{k}} G_{\omega+\nu, \bm{k}+\bm{q}}
R_{\omega\bm{k}} R_{\omega+\nu, \bm{k}+\bm{q}},
\end{align}
with 
$R_{\omega\bm{k}} = g_{\omega}^{-1} (\Delta_{\omega} - \epsilon_{\bm{k}})^{-1}$.
Equation~(\ref{eq:chi_lat}) reduces to the DMFT formula if the DMFT limit, $\tilde{\Sigma}_{\omega\bm{k}}=0$, is taken.

\begin{figure}[tb]
	\begin{center}
	\includegraphics[width=\linewidth]{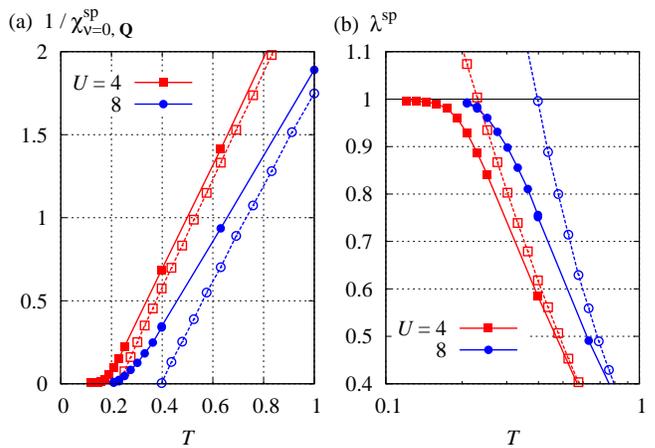}
	\end{center}
	\caption{(Color online) Temperature dependence of (a) the static AFM susceptibility $\chi^{\rm sp}_{\nu=0, \bm{Q}}$ and (b) the largest eigenvalue $\lambda^{\rm sp}$ of the matrix $\hat{A}$ at half-filling $n=1$. The closed symbols and open symbols show results in the present approximation and within DMFT, respectively.}
	\label{fig:T-afm}
\end{figure}

Figure~\ref{fig:T-afm}(a) shows the temperature dependence of the inverse of the static AFM susceptibility $\chi^{\rm sp}_{\nu=0, \bm{Q}}$, where $\bm{Q}=(\pi,\pi)$ is the nesting vector.
The DMFT result is also plotted for comparison.
From these data, we find that the susceptibility does not diverge in the ladder approximation, while the DMFT susceptibility diverges and obeys the Curie-Weiss law.

In the regime where fluctuations are strong, or more precisely, at $T \lesssim T_{\rm N}^{\rm DMFT}$ with $T_{\rm N}^{\rm DMFT}$ being the DMFT N\'eel temperature, the method presented in Appendix~\ref{app:stabilize} is essential to achieve convergence. Here we only mention that this method does not change the equations, and is simply a way of obtaining a converged solution.
A spurious divergence of $\chi^{\rm sp}$, which may arise during the iteration, is removed.
In this procedure, the main quantity we need to check is the dimensionless matrix $\hat{A}$ defined by
$(\hat{A})_{\omega \omega'} = \gamma^{\rm sp}_{\omega \omega'; 0} \tilde{\chi}^0_{\omega'; 0, \bm{Q}}$.
The condition for the divergence of $\Gamma^{\rm sp}$ in Eq.~(\ref{eq:Gamma}), and hence of $\chi^{\rm sp}$ is $\lambda^{\rm sp}=1$ where $\lambda^{\rm sp}$ denotes the largest eigenvalue of $\hat{A}$~\cite{footnote-lambda}.
The temperature dependence of $\lambda^{\rm sp}$ is shown in Fig.~\ref{fig:T-afm}(b).
It turns out that $\lambda^{\rm sp}$ approaches 1 with decreasing $T$ in the ladder approximation.

\begin{figure}[tb]
	\begin{center}
	\includegraphics[width=0.9\linewidth]{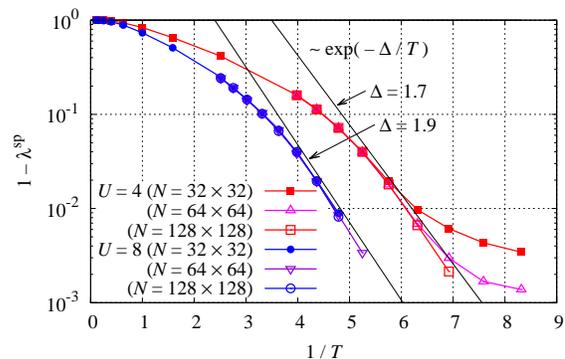}
	\end{center}
	\caption{(Color online) A scaling plot: $1-\lambda^{\rm sp}$ as a function of $1/T$. Results for different system sizes, $N=32\times32$, $64\times64$ and $128\times128$, are shown for comparison. The solid lines indicate the scaling $1-\lambda^{\rm sp} \propto \exp(-\Delta/T)$.}
	\label{fig:T-afm-scaling}
\end{figure}

In the critical regime, the susceptibility diverges exponentially toward $T=0$: $\chi \sim e^{\beta \Delta}$~\cite{Chakravarty88, Hasenfratz91}. It follows that $\lambda^{\rm sp}$ approaches 1 according to $1-\lambda^{\rm sp} \propto e^{-\beta \Delta}$.
In order to check this behavior, we plot $1-\lambda^{\rm sp}$ as a function of $1/T$ in Fig.~\ref{fig:T-afm-scaling}.
Results for larger system sizes, $N=64\times64$ and $128\times128$, are plotted as well.
It turns out that the data for different system sizes deviate from each other at low temperatures such that $1-\lambda^{\rm sp} \lesssim 10^{-2}$. It indicates that the slow decays for $N=32\times32$ and $64\times64$ observed at $1/T \gtrsim 7$ are artifacts due to a finite-size effect.
Apart from the finite-size effect, the results agree with the expected scaling $1-\lambda^{\rm sp} \propto e^{-\beta \Delta}$ indicated by the solid lines.
We thus conclude that our approximation correctly reproduces the N\'eel temperature of $T_{\rm N}=0$ required from the Mermin-Wagner theorem.

\section{Superconductivity}
\label{sec:SC}

\subsection{Formulas for pairing susceptibilities}
In this section, we discuss the superconductivity in the doped regime.
We first derive a formula for the pairing susceptibility of the dual fermions.
The susceptibilities of the dual fermions can be transformed to those of the original electrons~\cite{Li08, Brener08, Hafermann-book}.
Actually, numerical transformations cannot be performed in the case of unconventional (momentum-dependent) order parameters because the susceptibility matrix is too large to store in memory [see Eq.~(\ref{eq:BS-pp})].
However, since the diverging point is common to both susceptibilities, we can determine the transition temperature from the dual-fermion susceptibility without transforming to the electron susceptibility.

\begin{figure}[tb]
	\begin{center}
	\includegraphics[width=\linewidth]{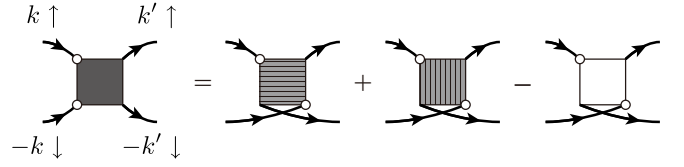}
	\end{center}
	\caption{The pairing interaction (the irreducible vertex for the pairing susceptibility) $\Gamma^{\rm pp}$ in the ladder approximation. The box with stripes stands for the renormalized vertex $\Gamma$ in Fig~\ref{fig:diagrams}(c).}
	\label{fig:pairing}
\end{figure}

We consider Cooper pairs with opposite spin directions of the constituent electrons.
With a form factor $\phi_{k}$ which depends on both $\bm{k}$ and $\omega$, 
the order parameter $\Phi$ is expressed as 
$\Phi = \sum_k \phi_{k} \langle f_{k\uparrow}f_{-k\downarrow} \rangle_{\tilde{\cal S}}$.
The static susceptibility for this pairing is defined by
$\sum_{kk'} \phi_{k} \tilde{P}_{kk'} \phi_{k'}^*$
where
\begin{align}
\tilde{P}_{kk'} 
= \langle f_{k\uparrow} f_{-k \downarrow} f^*_{-k'\downarrow} f^*_{k'\uparrow} \rangle_{\tilde{\cal S}}.
\end{align}
The Bethe-Salpeter equation for this Green's function is written as
\begin{align}
\tilde{P}_{kk'} = \tilde{P}^0_{k} \delta_{kk'}
- \frac{T}{N}\sum_{k''} \tilde{P}^0_{k} \Gamma^{\rm pp}_{kk''} \tilde{P}_{k''k'},
\label{eq:BS-pp}
\end{align}
where
\begin{align}
\tilde{P}^0_{k} = \tilde{G}_k \tilde{G}_{-k}.
\end{align}
For the irreducible vertex part $\Gamma^{\rm pp}$, we take account of effective interactions mediated by the spin and charge fluctuations. Hence, $\Gamma^{\rm pp}$ is given in terms of the renormalized vertex in Eq.~(\ref{eq:Gamma}) as~\cite{Hafermann-book,Hafermann09-2}
\begin{align}
\Gamma^{\rm pp}_{kk'}
= &-\Gamma^{\uparrow\downarrow\downarrow\uparrow}_{\omega, -\omega'; \omega'-\omega, \bm{k}'-\bm{k}}
+\Gamma^{\uparrow\downarrow\uparrow\downarrow}_{\omega, \omega'; -\omega-\omega', -\bm{k}-\bm{k}'}
\nonumber\\
&+ \gamma^{\uparrow\downarrow\downarrow\uparrow}_{\omega, -\omega'; \omega'-\omega}.
\label{eq:Gamma-pp}
\end{align}
The first term in Eq.~(\ref{eq:Gamma-pp}) incorporates the charge and longitudinal spin fluctuations, and the second term the transverse spin fluctuations. The third term subtracts their double counting.
A diagrammatic representation for $\Gamma^{\rm pp}$ is shown in Fig.~\ref{fig:pairing}.

Without magnetic field, 
the pairing susceptibility is classified according to the total spin of the pair.
For this purpose, we replace the pair operator by its symmetrized or anti-symmetrized form:
\begin{align}
f_{k\uparrow} f_{-k \downarrow} \to \frac{1}{\sqrt{2}} (f_{k\uparrow} f_{-k \downarrow} \mp f_{k\downarrow} f_{-k \uparrow}).
\end{align}
Here, $-$ corresponds to the spin singlet and $+$ to the spin triplet.
The corresponding pairing susceptibility is expressed as
\begin{align}
\tilde{P}_{k,k'}^{\pm} 
= \tilde{P}_{k,k'} \pm \tilde{P}_{k,-k'}.
\end{align}
Hence, the inversion of the fermionic frequency and momentum, $k=(\omega, \bm{k}) \to -k=(-\omega, -\bm{k})$, 
transforms $\tilde{P}_{kk'}^{\pm}$ as
$P^{\pm}_{k,k'} =\pm P^{\pm}_{k,-k'} =\pm P^{\pm}_{-k,k'} = P^{\pm}_{-k,-k'}$.
From Eq.~(\ref{eq:BS-pp}), we obtain the equation for $\tilde{P}_{kk'}^{\pm}$,
\begin{align}
\tilde{P}_{kk'}^{\pm} = \tilde{P}^0_{k} (\delta_{k,k'} \pm \delta_{k,-k'})
- \frac{T}{N}\sum_{k''} \tilde{P}^0_{k} \Gamma^{\rm pp\pm}_{kk''} \tilde{P}_{k''k'}^{\pm},
\label{eq:BS-pp-2}
\end{align}
where the (anti-)symmetrized vertex $\Gamma^{\rm pp\pm}_{kk'}$ is defined by
$\Gamma^{\rm pp\pm}_{kk'} = (\Gamma^{\rm pp}_{k,k'} \pm \Gamma^{\rm pp}_{k,-k'})/2$.
Their explicit expressions read
\begin{align}
\Gamma^{\rm pp+}_{kk'}
&= \frac{1}{4} \left[
(3\Gamma^{\rm sp}-\Gamma^{\rm ch})_{\omega, -\omega'; \omega'-\omega, \bm{k}'-\bm{k}}
- 2\gamma^{\rm sp}_{\omega, -\omega'; \omega'-\omega} \right]
\nonumber \\
&+ ( \omega' \to -\omega' ),
\\
\Gamma^{\rm pp-}_{kk'}
&= \frac{1}{4} \left[
-(\Gamma^{\rm sp}+\Gamma^{\rm ch})_{\omega, -\omega'; \omega'-\omega, \bm{k}'-\bm{k}}
+ 2\gamma^{\rm sp}_{\omega, -\omega'; \omega'-\omega} \right]
\nonumber \\
&- ( \omega' \to -\omega' ),
\end{align}
where $( \omega' \to -\omega' )$ is symbolic for the terms appearing before it with $\omega'$ replaced by $-\omega'$.

The dimension of the matrices is too large to solve Eq.~(\ref{eq:BS-pp-2}) numerically. We instead deal with an eigenvalue problem to determine the transition temperature and to extract the dominant pairing fluctuations.
Near the transition temperature, 
we may neglect the first term in Eq.~(\ref{eq:BS-pp-2}) to obtain the linear equation
\begin{align}
\label{eq:eigen-sc}
\hat{K}^{\pm} \phi = \lambda^{\rm SC} \phi,
\quad
(\hat{K}^{\pm})_{kk'}
= -\frac{T}{N} \tilde{P}^{0}_{k} \Gamma^{\rm pp\pm}_{kk'}.
\end{align}
We can demonstrate from the explicit form of $\Gamma^{{\rm pp}\pm}_{kk'}$ that the eigenvalues $\lambda^{\rm SC}$ are purely real. The condition for the divergence of the susceptibility is $\lambda^{\rm SC}_{\rm max}=1$ with $\lambda^{\rm SC}_{\rm max}$ being the largest eigenvalue.
The corresponding eigenfunction $\phi_k$ gives the form factor of the order parameter.

\subsection{Numerical results}

\begin{figure}[tb]
	\begin{center}
	\includegraphics[width=0.9\linewidth]{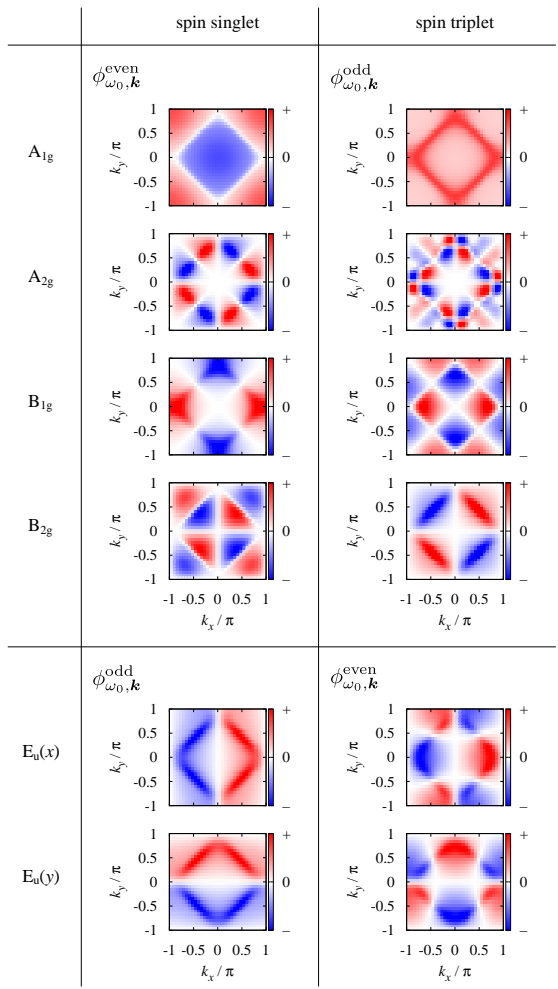}
	\end{center}
	\caption{(Color online) Momentum dependence of the eigenfunctions $\phi_{\omega_0,\bm{k}}$ of Eq.~(\ref{eq:eigen-sc}) for $U=8$, $\delta=0.14$ and $T=0.1$, where $\omega_0=\pi T$. 
	Either the even-frequency part $\phi^{\rm even}_{\omega_0, \bm{k}}$ or the odd-frequency part $\phi^{\rm odd}_{\omega_0, \bm{k}}$ is plotted depending on which is allowed by the Pauli principle.}
	\label{fig:pair-sym}
\end{figure}

We evaluated the largest eigenvalues $\lambda^{\rm SC}_{\rm max}$ of Eq.~(\ref{eq:eigen-sc}) by a kind of power method.
In this calculation, we enforced a particular spatial symmetry to pick up an eigenfunction belonging to a certain irreducible representation (see Appendix~\ref{app:pair_sym} for details).
In this way, we computed 10 types of pairings (2 spin symmetries $\times$ 5 spatial symmetries), which have the largest eigenvalue in each symmetry class.
The phase of the eigenfunction is arbitrary in the linear equation.
We determined the phase factor so that the component which has the largest absolute value becomes a real number.
Then, all components of $\phi_k$ become real. 
Finally, we define even- and odd-frequency parts,
$\phi_{\omega \bm{k}}^{\rm even} = \phi_{\omega \bm{k}} + \phi_{-\omega \bm{k}}$ and
$\phi_{\omega \bm{k}}^{\rm odd} = \phi_{\omega \bm{k}} - \phi_{-\omega \bm{k}}$,
to see the frequency dependence.
We have confirmed that either $\phi^{\rm even}$ or $\phi^{\rm odd}$ vanishes to fulfill the Pauli principle, e.g., $\phi^{\rm odd}=0$ for the spin-singlet with symmetry A$_{\rm 1g}$.

We first show eigenfunctions $\phi_{\omega \bm{k}}$ obtained in the way described above.
Figure~\ref{fig:pair-sym} shows the momentum dependence of $\phi_{\omega \bm{k}}$ with the lowest Matsubara frequency, $\omega_0=\pi T$.
The main feature is that some functions have only minimal nodes required from the symmetry and the rest have additional nodes. In the A$_{\rm 1g}$ symmetry, for example, there is no node for the triplet, while a line node exists on the Fermi level for the singlet (i.e., extended s-wave symmetry, $\cos k_x + \cos k_y$).

\begin{figure}[tb]
	\begin{center}
	\includegraphics[width=\linewidth]{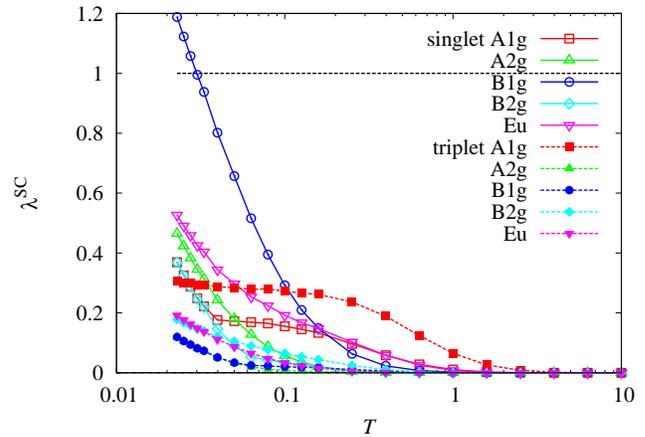}
	\end{center}
	\caption{(Color online) Temperature dependence of the eigenvalues $\lambda^{\rm SC}$ of Eq.~(\ref{eq:eigen-sc}) for $U=8$ and $\delta=0.14$.}
	\label{fig:lambda_sc}
\end{figure}

Which type of superconductivity actually occurs is examined from the temperature dependence of $\lambda^{\rm SC}$.
It can be seen from Fig.~\ref{fig:lambda_sc} that $\lambda^{\rm SC}$ for the spin-singlet B$_{\rm 1g}$ (d$_{x^2-y^2}$) symmetry crosses 1 as expected.
The transition temperature $T_{\rm c}$ is estimated to be $T_{\rm c} \simeq 0.030$ for these parameters. The doping dependence of $T_{\rm c}$ is plotted in the phase diagram in Fig.~\ref{fig:phase-dope}.

\section{Phase separation}
\label{sec:PS}
\begin{figure}[tb]
	\begin{center}
	\includegraphics[width=\linewidth]{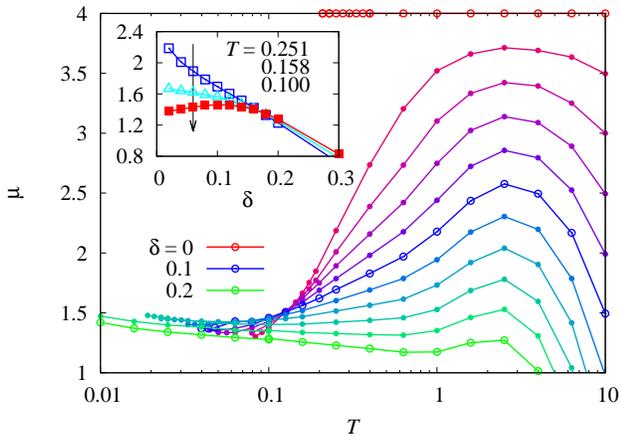}
	\end{center}
	\caption{(Color online) Temperature dependence of the chemical potential $\mu$. The doping $\delta$ is varied from 0 to 0.2 in 0.02 steps. The inset shows $\mu$ as a function of $\delta$ for fixed $T$.}
	\label{fig:mu}
\end{figure}

Our next interest lies in the paramagnetic state above $T_{\rm c}$ and near the Mott insulator.
In this regime, we found an instability of the uniform charge fluctuations.
Figure~\ref{fig:mu} shows the temperature dependence of the chemical potential $\mu$ for several values of doping $\delta=1-n$ for $U=8$. 
The decrease of $\mu$ below $T\simeq 1$ is due to the development of a Mott gap. At around $T=0.1$, some lines for different doping levels intersect.
It means that $\mu$ is a non-monotonic function of $\delta$ at low temperatures as shown in the inset of Fig.~\ref{fig:mu}. This behavior indicates a phase separation as explained below.

At $T=0.1$ in the inset of Fig.~\ref{fig:mu}, there exists two solutions with different doping, say $\delta_1$ and $\delta_2$. Actually, the Mott insulator with $\delta=0$ is also a solution in this case.
Hence, there are three solutions ($\delta_0=0<\delta_1<\delta_2$), two of which ($\delta_0$ and $\delta_2$) are thermodynamically stable and one ($\delta_1$) is unstable.
In order to make the average doping $\bar{\delta}$ at $0 < \bar{\delta} < \delta_2$, the system becomes spatially inhomogeneous between the Mott insulator with $\delta=0$ and the metallic state with $\delta=\delta_2$.

We define the temperature $T_{\rm PS}$ for the phase separation by the point where two lines intersect in Fig.~\ref{fig:mu}. 
It corresponds to the so-called spinodal point where the uniform charge susceptibility diverges.
The result for $T_{\rm PS}$ is plotted in the phase diagram of Fig.~\ref{fig:phase-dope}.
Below $T_{\rm PS}$, the homogeneous solution is thermodynamically unstable.

Before concluding the paragraph, we comment on a technical issue related to these observations.
The uniform charge susceptibility can be computed in two different ways: Either from the chemical potential as discussed above or from the correlation function as presented in Section~\ref{sec:AFM} for the spin channel. In the present approximation, the two results are not consistent. 
Indeed, we found no divergence of the uniform charge susceptibility computed from the correlation function.
In this case, the one computed from the chemical potential is more reliable in the sense that derivative of the self-energy with respect to $\mu$ is taken strictly, while the correlation function incorporates only a part of the corresponding diagrams.
To improve consistency, i.e., to obtain divergence in the correlation function, we need more elaborated treatment of the irreducible vertex to satisfy the Ward identity~\cite{Hafermann-arXiv}.

\section{Unconventional density waves}
\label{sec:DW}
\begin{figure}[tb]
	\begin{center}
	\includegraphics[scale=0.7]{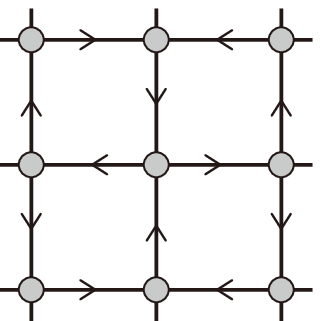}
	\end{center}
	\caption{The local-current configuration in the d-DW state.}
	\label{fig:stag_flux}
\end{figure}

In the previous section, we discussed the phase separation taking place near the Mott insulator. 
In this section, we examine the possibility of another phase, which has been discussed extensively, namely, the staggered flux state or the d-DW state~\cite{Chakravarty01, Kotliar88, Honerkamp02, Macridin04, Lu12}.
To make our formulation general, we consider both spin and charge channels ($\alpha=\text{sp}, \text{ch}$), arbitrary wave vectors $\bm{q}$, and arbitrary spatial symmetry.
The order parameter $\Psi^{\alpha\eta}_{\bm{q}}$ is defined by
\begin{align}
\Psi^{\alpha\eta}_{\bm{q}}
=\sum_{\omega\bm{k}\sigma}
\sigma^{\alpha}_{\sigma\sigma}
\psi_{\bm{k}}^{\eta}
\langle
f_{\omega\bm{k}\sigma}^{*}
f_{\omega\bm{k}+\bm{q}\sigma}
\rangle_{\tilde{\cal S}},
\end{align}
where $\sigma^{\rm ch}=\sigma^0$ and $\sigma^{\rm sp}=\sigma^z$.
The index $\eta$ labels different form factors $\psi_{\bm{k}}^{\eta}$.
The d-DW corresponds to $\Psi_{\bm{Q}}^{\rm ch, d}$ with the form factor
$\psi_{\bm{k}}^{\rm d}=i(\cos k_x - \cos k_y)$, while the ordinary DW is given by $\psi_{\bm{k}}^{\rm s}=1$.
In the real-space representation, the d-DW exhibits a local current 
$i\langle f_{\bm{r}}^* f_{\bm{r}+\bm{x}} \rangle - i\langle f_{\bm{r}+\bm{x}}^* f_{\bm{r}} \rangle$ 
which aligns as in Fig.~\ref{fig:stag_flux}.
Following the same reasoning as for the pairing correlations, we consider susceptibilities of the dual fermions.
The susceptibility corresponding to $\Psi^{\alpha\eta}_{\bm{q}}$ is given by
$\sum_{kk'} \psi_{k}^{\eta} \tilde{\chi}_{kk';q}^{\alpha} (\psi_{k'}^{\eta})^*$ with
\begin{align}
\tilde{\chi}_{kk';q}^{\alpha}
=
\frac{1}{2} \sum_{\sigma\sigma'}
\sigma^{\alpha}_{\sigma\sigma} \sigma^{\alpha}_{\sigma'\sigma'}
\langle f^*_{k\sigma} f_{k+q \sigma} f^*_{k'+q\sigma'} f_{k'\sigma'} \rangle_{\tilde{\cal S}}.
\end{align}
The susceptibility formula in Eq.~(\ref{eq:chi_lat}) does not give rise to the unconventional DW, since the irreducible vertex is local in this formula.
Higher-order processes need to be taken into account.

\begin{figure}[tb]
	\begin{center}
	\includegraphics[]{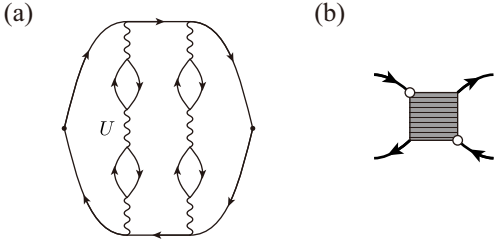}
	\end{center}
	\caption{(a) An exemplary susceptibility diagram which contributes to the unconventional CDW. (b) A diagram for the dual-fermion irreducible vertex $\Gamma'$ in Eqs.~(\ref{eq:vertex-dw-ch}) and (\ref{eq:vertex-dw-sp}). The box with stripes stands for the renormalized vertex $\Gamma$ in Fig~\ref{fig:diagrams}(c).}
	\label{fig:vertex-dw}
\end{figure}

In order to see which processes enhance unconventional DWs, it is instructive to analyze influences of intersite interactions $H_{\rm int}$ in the mean-field approximation or the random-phase approximation (RPA)~\cite{Ozaki92, Ikeda98}. 
We consider the nearest-neighbor repulsion $V$ and the AFM exchange interaction $J$:
\begin{align}
H_{\rm int}
= \frac{1}{2} \sum_{\langle i,j\rangle} \sum_{\sigma \sigma'}
\left[
V 
c_{i\sigma}^{\dag} c_{i\sigma}
c_{j\sigma'}^{\dag} c_{j\sigma'}
+J 
c_{i\sigma}^{\dag} c_{i\sigma'}
c_{j\sigma'}^{\dag} c_{j\sigma}
\right].
\end{align}
The staggered susceptibility $\chi^{\alpha\eta}_{\bm{Q}}$ in RPA is given by
\begin{align}
\chi^{\alpha\eta}_{\bm{Q}}=[(\chi^{0\eta}_{\bm{Q}})^{-1} - I_{\bm{Q}}^{\alpha\eta}]^{-1},
\end{align}
where
$\chi^{0\eta}_{\bm{Q}}=-(T/N)\sum_{\bm{k}} |\psi_{\bm{k}}^{\eta}|^2 G_{\omega\bm{k}} G_{\omega\bm{k}+\bm{Q}}$.
The sign of $I_{\bm{Q}}^{\alpha\eta}$ determines whether the fluctuations are enhanced ($I_{\bm{Q}}^{\alpha\eta}>0$) or suppressed ($I_{\bm{Q}}^{\alpha\eta}<0$).
Contributions to $I_{\bm{Q}}^{\alpha\eta}$ from $U$, $V$ and $J$ are summarized in Table~\ref{tab:RPA}~\cite{Ozaki92}. 
There are two types of diagrams in the RPA: The bubble-type (upper table row) and the ladder-type (lower). 
It turns out that the ladder-type diagram for $V$ and $J$ may cause an unconventional CDW and/or SDW~\cite{footnote-uDW}.
For the Hubbard model without $V$ and $J$, the local repulsion $U$ gives rise to AFM fluctuations, which effectively have an effect similar to that of the $J$-term. Hence, there is a chance that the unconventional CDW is induced by processes beyond RPA.
This spin-fluctuation mediated interaction is taken into account, for example, by the diagram in Fig.~\ref{fig:vertex-dw}(a).
This process is indeed included in the irreducible vertex derived by a functional derivative of the Luttinger-Ward functional in the FLEX~\cite{FLEX1}.
\begin{table}[t]
\begin{center}
\caption{Effect of different diagrams and interaction terms on the staggered susceptibilities in the RPA. The signs $+$ and $-$ indicate enhancement and suppression, respectively.}
\label{tab:RPA}
\begin{tabular}{c|ccc}
\hline
& $U$ & $V$ & $J$ \\
\hline
\parbox[c]{3.5cm}{\includegraphics[width=\linewidth]{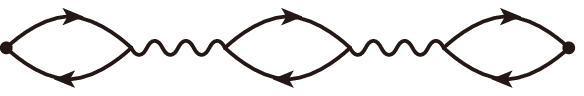}}
& \parbox[c]{40pt}{$+$SDW \\ $-$CDW} & $+$CDW & \parbox[c]{40pt}{$+$SDW \\ $+$CDW}
\\
\parbox[c]{3.5cm}{\includegraphics[width=\linewidth]{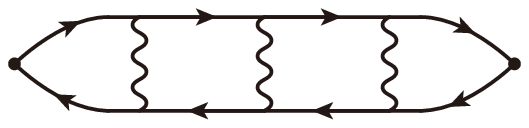}}
& $+$SDW & \parbox[c]{40pt}{$+$uSDW \\ $+$uCDW} & $+$uCDW
\\
\hline
\end{tabular}
\end{center}
\end{table}

In the dual-fermion approach, the effective interaction mediated by spin fluctuations can be constructed using the renormalized vertex $\Gamma$ in Eq.~(\ref{eq:Gamma}) and Fig.~\ref{fig:diagrams}(c). 
We note that $\Gamma$ contains both bubble-type and ladder-type diagrams and more if it is written with $U$, since the interaction vertex $\gamma$ is fully antisymmetrized.
Linearizing the Bethe-Salpeter equation for the static susceptibility as in the case of superconductivity, we obtain the eigenvalue equation
\begin{align}
\hat{L}^{\alpha}_{\bm{q}} \psi = \lambda^{\rm DW} \psi,
\quad
(\hat{L}^{\alpha}_{\bm{q}})_{kk'} = -\frac{T}{N} G_{\omega\bm{k}} G_{\omega,\bm{k}+\bm{q}}
\Gamma'^{\alpha}_{kk'}.
\label{eq:eigen-dw}
\end{align}
The irreducible vertex part $\Gamma'$ is given in terms of $\Gamma$ as
\begin{align}
\Gamma'^{\rm ch}_{kk'} 
&= -\frac{1}{2} ( 3 \Gamma^{\rm sp} + \Gamma^{\rm ch} )_{\omega, \omega; \omega'-\omega, \bm{k}'-\bm{k}},
\label{eq:vertex-dw-ch}
\\
\Gamma'^{\rm sp}_{kk'} 
&= \frac{1}{2} ( \Gamma^{\rm sp} - \Gamma^{\rm ch}  )_{\omega, \omega; \omega'-\omega, \bm{k}'-\bm{k}}.
\label{eq:vertex-dw-sp}
\end{align}
Figure~\ref{fig:vertex-dw}(b) shows a diagram corresponding to the vertex $\Gamma'$.
The matrix $\hat{L}$ in Eq.~(\ref{eq:eigen-dw}) is non-hermitian so that the eigenvalues $\lambda^{\rm DW}$ are complex numbers in general. By numerical calculations, we found that $\lambda^{\rm DW}$ consists of purely real numbers as well as complex numbers.
We have computed eigenvalues which has the largest real part by means of the Arnoldi method~\cite{Arpack}.
As in the case of superconductivity, we obtained 5 eigenvalues with different spatial symmetries for each channel $\alpha=\text{sp, ch}$.

\begin{figure}[tb]
	\begin{center}
	\includegraphics[width=\linewidth]{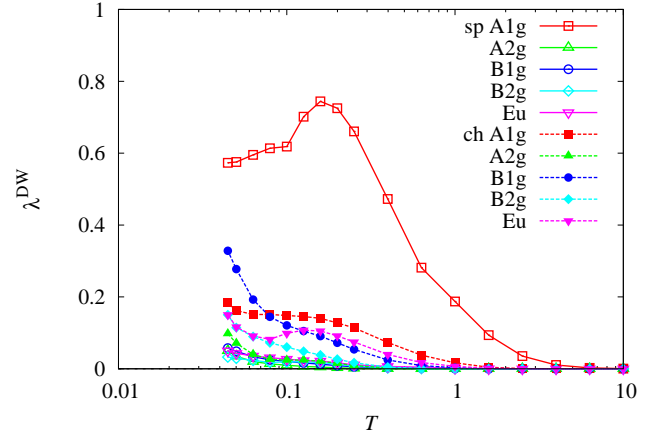}
	\end{center}
	\caption{(Color online) Temperature dependence of the eigenvalues $\lambda^{\rm DW}$ of Eq.~(\ref{eq:eigen-dw}) with $\bm{q}=\bm{Q}$. The parameters are $U=8$ and $\delta=0.08$.}
	\label{fig:T-dw}
\end{figure}

Figure~\ref{fig:T-dw} shows thus obtained eigenvalues for $\bm{q}=\bm{Q}$ as a function of temperature.
The largest fluctuation is the ordinary antiferromagnetism (sp-A$_{\rm 1g}$) as expected. The cusp around $T=0.1$ is due to a cross of eigenvalues between a purely real number on the high-$T$ side and a complex number on the low-$T$ side. On the other hand, at low temperatures, there is a significant enhancement of the leading eigenvalue corresponding to the eigenfunction with B$_{\rm 1g}$ (d$_{x^{2}-y^{2}}$) symmetry in the charge channel. This fluctuation corresponds to the staggered flux state or the d-DW state. This eigenvalue may exceed the one in the AFM spin channel at lower temperatures. An extrapolation of $\lambda^{\rm DW}$ to lower temperature indicates that the transition to the d-DW state takes place at $T\sim 0.01$.
However, the d-SC has a larger transition temperature as shown in Fig.~\ref{fig:phase-dope}. Therefore, the d-DW state actually does not realize in our approximation. We did not find any parameter regime where the d-DW state is superior to d-SC in the parameters we examined. 

\section{Discussions and Summary}
\label{sec:summary}

We have applied the dual-fermion approach to the two-dimensional Hubbard model. 
The AFM fluctuations have been taken into account by the ladder diagrams, which constitute the leading correction to the DMFT in terms of a $1/d$ expansion. 
Practically, there was a convergence issue which prevents calculations below the mean-field critical temperature.
By solving this technical problem, we are able to extend the applicability of the approach to significantly lower temperatures. 

Possible phase transitions have been investigated from susceptibilities in both the particle-hole and particle-particle channels.
For both calculations, we used a physically equivalent irreducible vertex representing spin-fluctuation mediated interactions.
Thus, we compared fluctuations of spin/charge DW and singlet/triplet superconductivity for all spatial symmetries including d-SC and d-DW.
The conclusion obtained is summarized in the phase diagrams in Section~\ref{sec:overview}.
The leading instability under doping is the d-SC as expected.
The d-DW fluctuations also show a clear tendency toward divergence at low temperatures.
However, the estimated transition temperature for the d-DW is below the $T_{\rm c}$ of d-SC, for all parameters considered. 
Our result hence supports the absence of a d-DW, which was yielded by several approximations~\cite{Honerkamp02, Macridin04, Lu12, Yokoyama-commun}.

At temperatures above $T_{\rm c}$ and low doping, we observed phase separation between the Mott insulator with $\delta=0$ and metallic region with $\delta\neq0$.
The existence of the phase separation agrees with other numerical calculations~\cite{Capone06,Macridin06,Aichhorn07, Yokoyama13, Misawa-arXiv}, but conflicts with QMC results~\cite{Moreo91,Becca00}.
Provided that the instability is not an artifact, a possible reason for the discrepancy is the system size: We employ $N=32 \times 32 = 1024$ lattice sites, while the QMC calculations were performed for smaller size, $N<200$, because of a sign problem.
The region of the phase separation extends to $\delta \simeq 0.15$ for $U=8$, and therefore a pure d-SC occurs only in the limited region $0.15 \lesssim \delta \lesssim 0.18$. 
This estimation is in quantitative agreement with the cluster DMFT~\cite{Capone06}.

The dual-fermion approach is complemental to the cluster DMFT among theories based on DMFT.
The ladder approximation in the dual fermion, in particular, aims at incorporating long-range fluctuations, while the cluster DMFT incorporates only short-range correlations.
Hence, it would be informative to summarize consistency and inconsistency between those results to clarify characteristics of two complemental approaches.
The instabilities reported from the cluster DMFT are consistent with ours: The phase separation as well as d-SC take place under doping~\cite{Capone06,Macridin06,Aichhorn07} and the d-DW is predominated by the d-SC~\cite{Macridin04}.
What can be reproduced by the dual-fermion approach but not by cluster DMFT is the critical behavior of the susceptibilities~\cite{Antipov14}, since a feedback of low-energy two-particle excitations to the self-energy is essential for it.
Our results for the AFM susceptibility exhibit a strong departure from the Curie-Weiss law, and are consistent with a critical temperature of $T_{\rm N}=0$ expected from the Mermin-Wagner theorem. 
This aspect will be considered in more detail elsewhere.

The short-range correlations, on the other hand, play an important role near the Mott insulator.
Its influence may arise in the doping dependence of $T_{\rm c}$. 
In our dual-fermion calculations, it turned out that $T_{\rm c}$ computed by neglecting the phase separation, namely, computed with the thermodynamically unstable solution, show no downturn as approaching the Mott insulator from finite doping.
In contrast, the cluster DMFT yields the dome shape of the d-SC phase~\cite{Lichtenstein00,Gull13}.

Further development beyond the dual-fermion approach has recently been attempted~\cite{Rubtsov12}.
The so-called dual boson theory introduces a bosonic counterpart of the dual fermion for the purpose of treating intersite interactions beyond mean-field theory~\cite{Loon-arXiv2} and collective excitations~\cite{Loon-arXiv1, Hafermann-arXiv}.
Furthermore, we expect that the dual boson in the spin channel yields formation of a intersite singlet, resulting in a reduction of $T_{\rm c}$ near the Mott insulator. 
This effect may be brought about by the coupling between spins and the vector bosonic field, which can be treated exactly by the recently developed algorithm~\cite{Otsuki13} based on the CT-QMC method.

\section*{Acknowledgments}
We acknowledge useful discussions with Y. Kuramoto, H. Yokoyama, H. Tsunetsugu, N. Tsuji, and M. Kitatani.
A part of the computations was performed in the ISSP Supercomputer Center, the University of Tokyo.
Two of us (JO and HH) acknowledge hospitality of the ISSP during the NHSCP2014 workshop.
This work was supported by JSPS KAKENHI Grant Number 26800172.
HH acknowledges support from the FP7/ERC, under Grant Agreement No. 278472-MottMetals.

\appendix

\section{Stabilization of self-energy calculation}
\label{app:stabilize}
In two-dimensional systems, the AFM susceptibility diverges exponentially for $T$ approaching zero~\cite{Chakravarty88, Hasenfratz91}. Since the ladder approximation explicitly takes the AFM fluctuations into account in the self-energy, the critically large fluctuations complicate the convergence of the self-energy iterations. In this Appendix, we present how to relieve this difficulty to get better convergence.

To show our idea of how to avoid the instability in the critical regime, we begin with the FLEX equations, which encounter the same problem in a simpler form. In FLEX, the susceptibility $\chi(q)$ is given by
\begin{align}
\chi(q) = \frac{\chi_0(q)}{1-U\chi_0(q)}.
\end{align}
Here, $\chi_0(q)$ is computed with the dressed Green's function $G(k)$, which is determined self-consistently.
Even though $\chi(q)$ is positive and finite in the converged solution, the right-hand side may diverge or become negative during the iteration if a trial $G(k)$ is not sufficiently close to the solution.
The self-energy evaluated from this susceptibility is divergent. This is the source of instability of the iteration.

To avoid this instability, we manipulate $\chi_0(q)$ so that $\chi(q)$ does not diverge.
Specifically, we replace $\chi_0(q)$ with
\begin{align}
\chi_0'(q) = \min[ \chi_0(q), (1-\eta)/U ],
\end{align}
where $\eta$ is a small constant.
If $\eta$ is too small, say $\eta=10^{-4}$, the instability may not be taken away.
Empirically, the FLEX iteration is stable down to $\eta \simeq 10^{-3}$.
We note that this replacement is done only to avoid the instability and to approach the solution. 
If the trial $G(k)$ is sufficiently close to the actual solution, this replacement is no longer necessary. The converged solution therefore is well defined.
In order to assure this, we should check whether the condition $U\chi_0<1-\eta$ is satisfied after the iteration is converged.
If this is not the case, it means that the actual solution of the equation is in the region $U\chi_0>1-\eta$.

We apply the above trick to the dual self-energy in Eq.~(\ref{eq:self-ladder}). 
Introducing a matrix notation for the fermionic frequencies, 
$(\hat{\Gamma}_{\nu\bm{q}})_{\omega\omega'} \equiv \Gamma_{\omega\omega'; \nu\bm{q}}$, 
$(\hat{\chi}^0_{\nu\bm{q}})_{\omega\omega'} \equiv \tilde{\chi}^0_{\omega; \nu\bm{q}} \delta_{\omega \omega'}$
and 
$(\hat{V}^{\alpha}_{\nu\bm{q}})_{\omega\omega'} \equiv V^{\alpha}_{\omega\omega'; \nu\bm{q}}$,
and omitting indices $\alpha$, $\nu$ and $\bm{q}$ for simplicity, 
the renormalized vertex $\Gamma$ in Eq.~(\ref{eq:Gamma}) and 
the effective interaction $V$ in Eq.~(\ref{eq:V_phi}) are rewritten in simpler forms
\begin{align}
\hat{\Gamma}
&= \hat{\gamma}
+T \hat{\gamma} \hat{\chi}^0 \hat{\Gamma},
\\
\hat{V} 
&= T \hat{\gamma} \hat{\chi}^0 [ 2\hat{\Gamma}-\hat{\gamma} ].
\end{align}
We diagonalize the dimensionless matrix $(T \hat{\gamma} \hat{\chi}^0)$ according to
\begin{align}
\hat{U}^{-1} (T \hat{\gamma} \hat{\chi}^0) \hat{U} = \lambda.
\end{align}
We note that the eigenvalues $\lambda_i$ are complex in general.
Using the diagonal matrix $\lambda$, the vertex $\hat{\Gamma}$ and $\hat{V}$ are expressed as
\begin{align}
\hat{\Gamma} 
&= \hat{U} (1-\lambda)^{-1} \hat{U}^{-1} \hat{\gamma},
\\
\hat{V} 
&= \hat{U} \lambda (1-\lambda)^{-1} (1+\lambda) \hat{U}^{-1} \hat{\gamma}.
\end{align}
It is clear from these expressions that the iteration becomes unstable once one of the eigenvalues $\lambda_i$ exceeds 1 (or more precisely, ${\rm Re} \lambda_i > 1$).
To avoid this instability, we replace ${\rm Re} \lambda_i$ with
\begin{align}
{\rm Re} \lambda'_i = \min({\rm Re} \lambda_i, 1-\eta),
\end{align}
during the iterations.
We remind that one needs to verify that the condition ${\rm Re} \lambda_i<1-\eta$ is fulfilled after convergence is reached.

\section{Spatial symmetry of pairing correlations}
\label{app:pair_sym}
In this Appendix, we present how to obtain eigenvectors with specific symmetry in Eqs.~(\ref{eq:eigen-sc}) and (\ref{eq:eigen-dw}).
In a power method and related algorithms, one computes a matrix-vector product $\hat{K} \phi^{\rm (old)}$ to obtain a new vector $\phi^{\rm (new)}$.
At this point, we restrict $\phi^{\rm (new)}$ to a subspace defined by the projection operator ${\cal P}$:
\begin{align}
\phi^{\rm (new)} = {\cal P} \hat{K} \phi^{\rm (old)}.
\end{align}
Thus, eigenvectors in the subspace are selectively computed.

\begin{table}[t]
\begin{center}
\caption{The character table for the point group D$_{4}$~\cite{grouptheory}. For the two-dimensional representation E$_u$, the diagonal element of the representation matrix is shown. The operation $C_4$ denotes $\pi/2$ rotation around the $z$ axis, and $C_2'$ and $C_2''$ denote $\pi$ rotations around the $x$ axis and the line $x=y$, respectively.}
\label{tab:D4}
\begin{tabular}{c|rrrrr}
& $E$ & $C_4$ & $C_4^2$ & $C_2'$ & $C_2''$  \\
\hline
A$_{\rm 1g}$ & 1 & 1 & 1 & 1 & 1 \\
A$_{\rm 2g}$ & 1 & 1 & 1 & $-1$ & $-1$ \\
B$_{\rm 1g}$ & 1 & $-1$ & 1 & 1 & $-1$ \\
B$_{\rm 2g}$ & 1 & $-1$ & 1 & $-1$ & 1 \\
E$_{\rm u}(x)$ & 1 & 0 & $-1$ & 1 & 0 \\
E$_{\rm u}(y)$ & 1 & 0 & $-1$ & $-1$ & 0
\end{tabular}
\end{center}
\end{table}

We consider an explicit expression for the projection operator ${\cal P}$.
In the square lattice, engenvectors $\phi$ belong to one of the irreducible representations $D$ in the point group D$_4$.
The symmetry property of $D$ is summarized in the character table in Table~\ref{tab:D4}.
There are 5 irreducible representations, 
$D={\rm A}_{\rm 1g}, {\rm A}_{\rm 2g}, {\rm B}_{\rm 1g}, {\rm B}_{\rm 2g}, E_{\rm u}$,
and 5 types of symmetry operations, 
${\cal C} = E, C_4, C_4^2, C_2', C_2''$. 
The value $\sigma=+1$ or $-1$ in the table shows the eigenvalue of the operation ${\cal C}$,
i.e., ${\cal C} \phi = \sigma \phi$, 
while $\sigma=0$ means that the operation changes the basis [e.g., $\pi/2$ rotation $C_4$ transforms E$_{\rm u}(x)$ to E$_{\rm u}(y)$]. 

We can project an arbitrary vector $\phi$ onto the irreducible representation $D$ by enforcing the symmetry property given in Table~\ref{tab:D4}.
Hence, the projection operator ${\cal P}(D)$ may be decomposed into a product
\begin{align}
{\cal P}(D) = \prod_{\cal C} {\cal Q}({\cal C}, \sigma({\cal C}, D)).
\end{align}
The operator ${\cal Q}({\cal C}, \sigma)$ singles out vectors which have the eigenvalue $\sigma$ of the operation ${\cal C}$. 
Since the eigenvalue $\sigma$ is either $+1$ or $-1$, we can implement the operation ${\cal Q}({\cal C}, \sigma) \phi_{\bm{k}} $ by
\begin{align}
{\cal Q}({\cal C}, \sigma) \phi_{\bm{k}} 
= \frac{1}{1+|\sigma|} ( \phi_{\bm{k}} + \sigma {\cal C}\phi_{\bm{k}} ).
\end{align}
We note that this operator is the identity when $\sigma=0$.


\begin{thebibliography}{99}
\bibitem{footnote-lambda}
We note that not all the eigenvalues are real numbers. 
But, we numerically confirmed that only pure real eigenvalues get close to 1.
\bibitem{footnote-uDW}
Which type of unconventional DW is favored depends on the range of interactions.
In the case of the nearest-neighbor interactions, fluctuations of d$_{x^2-y^2}$ symmetry $\psi_{\bm{k}}^{\rm d}$, the extended-s-wave $\psi_{\bm{k}}^{\rm ex-s}=\cos k_x + \cos k_y$, and p-wave symmetry $\psi_{\bm{k}}^{\rm p}=\sin k_x,\ \sin k_y$ are equally enhanced~\cite{Ozaki92}.
\bibitem{Arpack} We used ARPACK library, which provides calculations of eigenvalues and eigenfunctions for large matrices. We can selectively compute the eigenvalue which is the largest/smallest in the real part, in the imaginary part, or in the absolute value.
\end{thebibliography}
\end{document}